\documentclass[a4paper,10pt]{article}
\usepackage{pos}

\title{GeV Gamma-ray Counterparts of New Candidate Radio Supernova Remnants Reported in the GLEAM Survey}
\ShortTitle{GeV Gamma-ray Counterparts of New Radio Supernova Remnants Reported in the GLEAM Survey}

\author*[a]{B.M. Meşe}
\author[b]{T. Ergin}
\affiliation[a]{Department of Physics, Middle East Technical University, 
06800, Ankara, Turkey.}
\affiliation[b]{Middle East Technical University, Northern Cyprus Campus, 99738 Kalkanli via Mersin 10, Turkey.}
\emailAdd{mina.mese@metu.edu.tr}
\emailAdd{etulun@metu.edu.tr}

\abstract{
Recently the Galactic and Extra-galactic All-sky Murchison Widefield Array survey has published 27 new candidate radio supernova remnants (SNRs) which are located within the longitude ranges of  345° < l < 60° and 180° < l < 240°. To search for the gamma-ray counterparts of these candidate radio SNRs, we analyzed 14 years of {\it Fermi}-LAT data in the energy range of  1 - 300 GeV. There are three promising SNRs; G18.9$-$1.2, G23.1$+$0.1, and G28.3$+$0.2, which we detected at a significance level of $\sim$9$\sigma$, $\sim$13$\sigma$, and $\sim$12$\sigma$, respectively. Here we report the results of our morphological and spectral analyses of G18.9$-$1.2, G23.1$+$0.1, and G28.3$+$0.2. No extended gamma-ray emission is detected for any of these SNRs. Our analysis of the 3 SNRs' {\it Fermi}-LAT gamma-ray emission showed that their best-fit positions (if assumed point-like) overlap with the locations of the corresponding GLEAM counterparts.}

\FullConference{%
  7th Heidelberg International Symposium on High-Energy Gamma-Ray Astronomy (Gamma2022)\\
  4-8 July 2022\\
  Barcelona, Spain\\}
  
\begin{document}

\maketitle\section{Introduction}
\label{sec_intro}
\vspace{-0.2cm}
The number of supernova remnants (SNRs) in the Galaxy is estimated to be about 1000 - 2000 \citep{Li1991} \citep{Tammann1994}, while the number of observed SNRs is only a few hundred. This gap is probably caused by observational inefficiencies. Statistically, most SNRs are discovered at the radio wavelengths, which makes radio surveys crucial for new SNR detections. Murchison Widefield Array (MWA) telescope is a radio telescope located in Western Australia. It observes at 80 - 300 MHz band with a wide field of view. The detection of the SNRs is done with the Galactic and Extra-galactic All-sky MWA (GLEAM) data. A blind study was made (to avoid any possible bias), and two papers were published from this effort using the GLEAM survey’s Galactic data release to detect candidate radio SNRs located within the longitude ranges of 345$^{\circ}$ < l < 60$^{\circ}$ and 180$^{\circ}$ < l < 240$^{\circ}$: 1) \citet{Hurley-Walker2019a} detected 101 candidate radio SNRs 2) The more recent paper \citep{Hurley-Walker2019b} presented 27 new candidate radio SNRs. In this study, we analyzed the 27 SNRs listed in the second GLEAM paper in the gamma-ray energy band. GeV gamma rays are important messengers related to the production and acceleration processes of cosmic rays (CRs) happening within SNRs. GeV gamma-rays can be best observed from space telescopes, such as the {\it Fermi} Gamma-Ray Space Telescope. {\it Fermi}-LAT detects gamma-rays at energies up to $\sim$1 TeV. The angular resolution for 1 - 100 GeV is between 0.8$^{\circ}$ to 0.1$^{\circ}$ of 68\% (PSF) containment angle. Its energy resolution is smaller than 10\% for energies higher than 1 GeV\footnote{\url{https://www.slac.stanford.edu/exp/glast/groups/canda/lat_Performance.htm}}

{\bf G18.9$-$1.2}: This SNR was first discovered as G18.95$-$1.1 at the radio wavelengths \citep{Fürst1985}. The size of this SNR was measured to be 68$'\times\!~$ 60$'$ \citep{Hurley-Walker2019b}. Its age is about 2800 - 6100 years, and its distance ranges between 1.6 and 2.5 kpcs \citep{Ranasinghe2019}. In X-rays, it was discovered by ROSAT \citep{Aschenbach1991}. In gamma-rays, it was detected by {\it Fermi}-LAT and named 3FGL J1829.7-1304 \citep{Acero2015}. It is a shell-type SNR associated with SNR G018.9$-$01.1. The location of this source overlaps with 4FGL J1829$-$1256. In this region, a pulsar wind nebula (PWN) is thought to be associated with 4FGL J1829$-$1256, and there is a pulsar candidate CXOU J182913.1$-$125113. Also, there is a possible association with a molecular cloud (MC) \citep{Hewitt2009} \citep{Traverso1999}. Although Chandra, ROSAT, and ASCA telescopes detected this source as an extended source, {\it Fermi}-LAT could still not detect any extended gamma-ray emission.

{\bf G23.1$+$0.1:} Both the major and minor axes of the radio extension ellipse of this SNR are 26$'$ \citep{Hurley-Walker2019b}. It was detected by the High Energy Spectroscopic System (H.E.S.S.) at TeV gamma-ray energies and named HESS J1832$-$085 \citep{Abdalla2018}. Gamma-ray emission was reported in the region where G23.1$+$0.1 and HESS J1832$-$085 sources partially overlap \citep{Ergin2021}. The location of G23.1$+$0.1 is also coincident with that of 4FGL J1832.4$-$0847.

{\bf G28.3$+$0.2:} Each of the major and minor axes of the radio extension ellipse of this SNR was measured to be 14$'$ \citep{Hurley-Walker2019b}. G28.3$+$0.2 has four nearby pulsars. PSR J1841$-$0345 which is $\sim$56,000 years old, is the youngest one \citep{Morris2002}. Although this SNR might be associated with one of these pulsars, as it is not certain, the SNR's age and distance are not determined \citep{Hurley-Walker2019b}. The location of this shell-type SNR overlaps with 4FGL J1842.5$-$0359 and with an extra-galactic source, NVSS J184240$-$035858. However, no relationship was found between G28.3$+$0.2 and NVSS J184240$-$035858 \citep{Condon1998}. 
\vspace{-0.2cm}
\begin{table}
\caption{{\footnotesize The 27 SNRs listed are the sources at the \citet{Hurley-Walker2019b} paper. The Galactic longitude (l), Galactic latitude (b), Right Ascension (R.A.), and Declination (Dec.)} values -which are all in degrees- are taken from that paper. Test Statistic (TS) values are calculated assuming a {\it PowerLaw} point source model. Only the SNRs with '*' sign are calculated assuming a {\it LogParabola} point source model.}
\centering
\begin{tabular}{lccccccc} 
\hline
 \# & SNR Name & l[$^{\circ}$] & b[$^{\circ}$] & R.A.[$^{\circ}$] & Dec.[$^{\circ}$] & TS  \\
\hline
1 & G0.1-9.7 & 0.15 & -9.72 & 276.458 & -33.500 & 2.21 \\
2 & G2.1+2.7 & 2.16 & 2.76 & 265.042 & -25.650 & 3.45 \\
3 & G7.4+0.3 & 7.44 & 0.34 & 270.275 & -22.350 & 2.13 \\
4* & G18.9$-$1.2 & 18.95 & -1.25 & 277.517 & -13.000 & 92.92 \\
5 & G19.1-3.1 & 19.15 & -3.13 & 279.329 & -13.683 & -0.01 \\
6 & G19.7-0.7 & 19.74 & -0.71 & 277.396 & -12.050 & 1.44 \\
7 & G20.1-0.2 & 20.18 & -0.25 & 277.196 & -11.450 & 0.06 \\
8 & G21.8+0.2 & 21.82 & 0.20 & 277.563 & -9.783 & 0.95 \\
9 & G23.1$+$0.1 & 23.12 & 0.19 & 278.179 & -8.633 & 174.41 \\
10 & G24.0-0.3 & 24.09 & -0.34 & 279.108 & -8.017 & 0.00 \\
12* & G28.3$+$0.2 & 28.37 & 0.21 & 280.592 & -3.967 & 145.98 \\
13 & G28.7-0.4 & 28.78 & -0.46 & 281.375 & -3.900 & -0.01 \\
14 & G35.3-0.0 & 35.37 & -0.04 & 284.008 & 2.150 & 6.54 \\
15 & G230.4+1.2 & 230.46 & 1.30 & 112.238 & -14.933 & 1.24 \\
16 & G232.1+2.0 & 232.16 & 2.06 & 113.783 & -16.050 & 0.00 \\
17 & G349.1-0.8 & 349.12 & -0.83 & 260.100 & -38.517 & 06.01 \\
18 & G350.7+0.6 & 350.77 & 0.69 & 259.721 & -36.283 & -0.00 \\
19 & G350.8+5.0 & 350.82 & 5.05 & 255.467 & -33.667 & 0.02 \\
20 & G351.0-0.6 & 351.05 & -0.64 & 261.279 & -36.817 & 7.53 \\
21 & G351.4+0.4 & 351.41 & 0.48 & 260.379 & -35.883 & -0.00 \\
22 & G351.4+0.2 & 351.47 & 0.22 & 260.688 & -35.983 & -0.00 \\
23 & G351.9+0.1 & 351.90 & 0.15 & 261.058 & -35.667 & -0.00 \\
24 & G353.0+0.8 & 353.05 & 0.80 & 261.192 & -34.350 & 4.74 \\
25 & G355.4+2.7 & 355.44 & 2.76 & 260.867 &-31.267 & 0.06 \\
26 & G356.5-1.9 & 356.54 & -1.94 & 266.229 & -32.900 & 3.06 \\
27 & G358.3-0.7 & 358.38 & -0.77 & 266.192 & -30.717 & 14.90 \\
\hline
\end{tabular}
\end{table}
\section{Data Reduction and Analysis}
\vspace{-0.2cm}
\subsection{Data Reduction and Background Modelling}
\vspace{-0.1cm}
We used \texttt{fermitools}\footnote{\url{http://fermi.gsfc.nasa.gov/ssc/data/analysis/software}} version 1.2.23 and \texttt{fermipy}\footnote{\url{http://fermipy.readthedocs.io/en/latest/index.html}} version 0.17.4 analysis packages provided by NASA for the data reduction. We have used about 14 years of {\it Fermi}-LAT data from 04 July 2008 to varying dates from 14 December 2021 to 29 April 2022 for all sources listed in Table 1. For G18.9$-$1.2 and G28.3$+$0.2, the data set used is from 04 July 2008 to 29 April 2022, and for G23.1$+$0.1, it is from 04 July 2008 to 06 February 2022. The energy interval is 1 - 300 GeV and the Region of Interest (ROI) is a 15$^{\circ}$ circular area with the location of each source at the ROI's center. We used the {\tt Python} version of 2.7.14 and the {\tt ipython} version of 5.8.0. To visualize the gamma-ray sky maps, the image display and visualization program {\tt SAOImageDS9} was used. The most recent diffuse background models available, consisting of the galactic interstellar emission model \texttt{gll$\_$iem$\_$v07.fits} and the isotropic template model \texttt{iso$\_$P8R3$\_$SOURCE$\_$V2$\_$v1.txt}, were used to model the gamma-ray background emission. The 4FGL 8-year source catalog{\footnote{https://fermi.gsfc.nasa.gov/ssc/data/access/lat/8yr$\_$catalog/}} (\texttt{gll$\_$psc$\_$v22.fit}) and the {\tt make4FGLxml.py} script were downloaded from the {\it Fermi}-LAT web site and used to create the XML\footnote{extensible markup language} ROI model file fitted to the data. The extended source template archive (\texttt{LAT$\_$extended$\_$sources$\_$8years.tgz}) and the region files that contain the information on the sources' names, positions, and position errors were downloaded from the {\it Fermi}-LAT website and were used for visualizing the 4FGL catalog sources inside {\tt SAOImageDS9}.
\vspace{-0.2cm}
\subsection{TS Map and Extension Tests}
\vspace{-0.2cm}
For the analysis of all 27 SNRs listed in Table 1, the energy interval is chosen to be 1 - 300 GeV. Each of these SNRs is treated as a point-like gamma-ray source with a {\it PowerLaw} type spectrum. Those SNRs with a final TS value of 25 or higher are selected for further analysis. As the next step, the extension tests are conducted and binned analysis with energy dispersion correction is performed, using all events (evtype=3).

TS maps are produced for G18.9$-$1.2, G23.1$+$0.1, and G28.3$+$0.2 with a binning value of 0$^{\circ}\!$.01 $\times$ 0$^{\circ}\!$.01. The number of energy bins per decade is 4 for all analyses performed. The time interval of the data sets are 04.08.2008 to 29.04.2022 for G18.9$-$1.2, 04.08.2008 to 06.02.2022 for G23.1$+$0.1, 04.08.2022 to 29.04.2022 for G28.3$+$0.2. The locations of G18.9$-$1.2, G23.1$+$0.1, and G28.3$+$0.2 in equatorial coordinates in degrees are as follows respectively: (277.517,$-$13), (278.179,$-$8.63) and (280.592,$-$3.96). In order to find the best-fit location of the gamma-ray emission, the {\tt localize} script was used. After deleting the associated source from the background model, it is re-added to the model at this best-fit location as a point source with the same spatial model given in the 4FGL source catalog. Then, using the {\tt extension} method, two 2D extension models were tested: {\it RadialGaussian} and {\it RadialDisk}. The TS$_{\rm ext}$\footnote{ TS$_{\rm ext}$ corresponds to the TS of the extension parameter, which is the division of the probability that the source is a point source ($L_{\rm pt}$) to the probability that the source is an extended source ($L_{\rm ext}$): TS$_{\rm ext}$ = -2log($L_{\rm ext}$/$L_{\rm pt}$). To detect a source as an extended gamma-ray source, the TS$_{\rm ext}$ value must be equal or higher than 25.} parameter resulting from the extension method quantifies the likelihood of the extended model compared to that of the point source hypothesis (through a likelihood ratio test). TS$_{\rm ext}$ = 25 is usually considered the threshold to claim significant evidence of extension. During the extension tests, the parameters associated with any 4FGL source related to one of the three selected SNRs (i.e. $N_0$, $\alpha$, and $\beta$, see Section \ref{section_sed}) were set free. After the extension tests, these 4GFL sources were deleted from the background model and added back, this time as point sources at their best-fit locations and with the same spatial models that were given in the 4FGL source catalog. TS maps were created using the corresponding best-fit parameters and shown in Figures 1, 2, and 3 for G18.9$-$1.2, G23.1$+$0.1, and G28.3$+$0.2, respectively.
\begin{figure*}
\centering 
\includegraphics[width=0.8\textwidth]{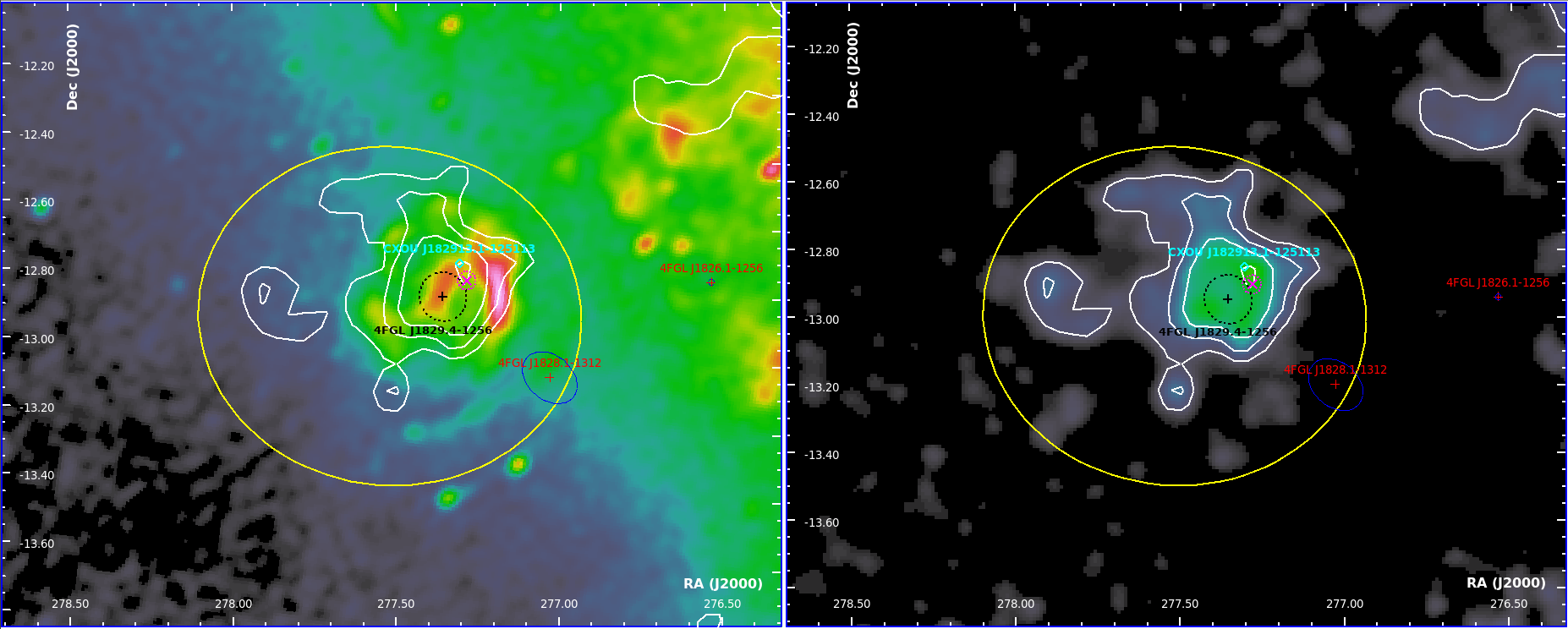}
\vspace{-0.2cm}
\caption{\footnotesize{Radio continuum of the ROI is shown at the left panel and the TS map of {\it Fermi}-LAT data at 1 - 300 GeV is shown at the right panel. The GLEAM location of G18.9$-$1.2 is at the center of the ROI. The white lines are the gamma-ray contours that are at TS values 4, 9, 16, and 25. The yellow ellipse is the GLEAM extension of G18.9$-$1.2. At the right panel, 4FGL J1829.4$-$1256 is deleted from the background model. Its location is shown with a black cross, while its error ellipse is shown with a black dashed ellipse. Other 4FGL sources are shown with red crosses and their error ellipses are shown with blue ellipses. The error ellipses of 4FGL sources are for 95\% confidence level.} The magenta X sign shows the best-fit location and the magenta dashed ellipse shows its error. CXOU J182913.1$-$125113 shown with cyan diamond is thought to be a pulsar associated with this SNR.}
\vspace{-0.2cm}
\label{figure1}
\end{figure*}
\vspace{-0.2cm}
\subsection{SED production}
\label{section_sed}
\vspace{-0.2cm}
The {\it LogParabola}\footnote{\url{https://fermi.gsfc.nasa.gov/ssc/data/analysis/scitools/source_models.html\#LogParabola}} spectral model was used for both G18.9$-$1.2 and G28.9+0.2, and the {\it PowerLaw}\footnote{\url{https://fermi.gsfc.nasa.gov/ssc/data/analysis/scitools/source_models.html\#PowerLaw}} spectral model was used for G23.1$+$0.1. In these models, $N_0$ is the normalization parameter. The spectral indices $\alpha$ and $\beta$ are used in the {\it LogParabola} model, while $\gamma$ is the spectral index of the {\it PowerLaw} model. The spectral analysis was done for the energy interval of 0.2 - 300 GeV and for each SNR the corresponding spatial model was chosen to be a point source model during the spectral analysis. The resulting spectral energy distribution (SED) graphs are shown on the left, middle, and right panels of Figure 4 for G18.9$-$1.2, G23.1$+$0.1, and G28.3$+$0.2, respectively.

\section{Results \& Conclusion}
\vspace{-0.2cm}
In this study, we analyzed 14 years of {\it Fermi}-LAT data of 27 SNR candidates of the GLEAM survey. As a result, we firmly detected 3 SNRs; G18.9$-$1.2, G23.1$+$0.1, and G28.3$+$0.2 in the gamma-ray energy range of 1 - 300 GeV with TS values being equal to or higher than 25 and 24 of these SNRs could not be detected in gamma-rays (i.e. TS < 25). The detected SNRs were analyzed further to find their best-fit positions and gamma-ray extension models in the 1 - 300 GeV energy band. By assuming that each SNR is a point source and using the spectral model of {\it LogParabola} for G18.9$-$1.2 and G28.9+0.2, and the {\it PowerLaw} model for G23.1$+$0.1, we computed the best-fit positions to be (R.A., Dec.) = (277$^{\circ}\!$.28 $\pm$ 0$^{\circ}\!$.03, -12$^{\circ}\!$.91 $\pm$ 0$^{\circ}\!$.03), (278$^{\circ}\!$.10 $\pm$ 0$^{\circ}\!$.01, -8$^{\circ}\!$.75 $\pm$ 0$^{\circ}\!$.02), and (280$^{\circ}\!$.64 $\pm$ 0$^{\circ}\!$.04, -4$^{\circ}\!$.07 $\pm$ 0$^{\circ}\!$.05) as given with the statistical error bars for the positions. The detection significance (approximated as $\sqrt{TS}$0) at the best-fit positions were found to be $\sim$9$\sigma$, $\sim$13$\sigma$, and $\sim$12$\sigma$ for G18.9$-$1.2, G23.1$+$0.1, and G28.3$+$0.2, respectively.

\begin{figure*}
\centering 
\includegraphics[width=0.8\textwidth]{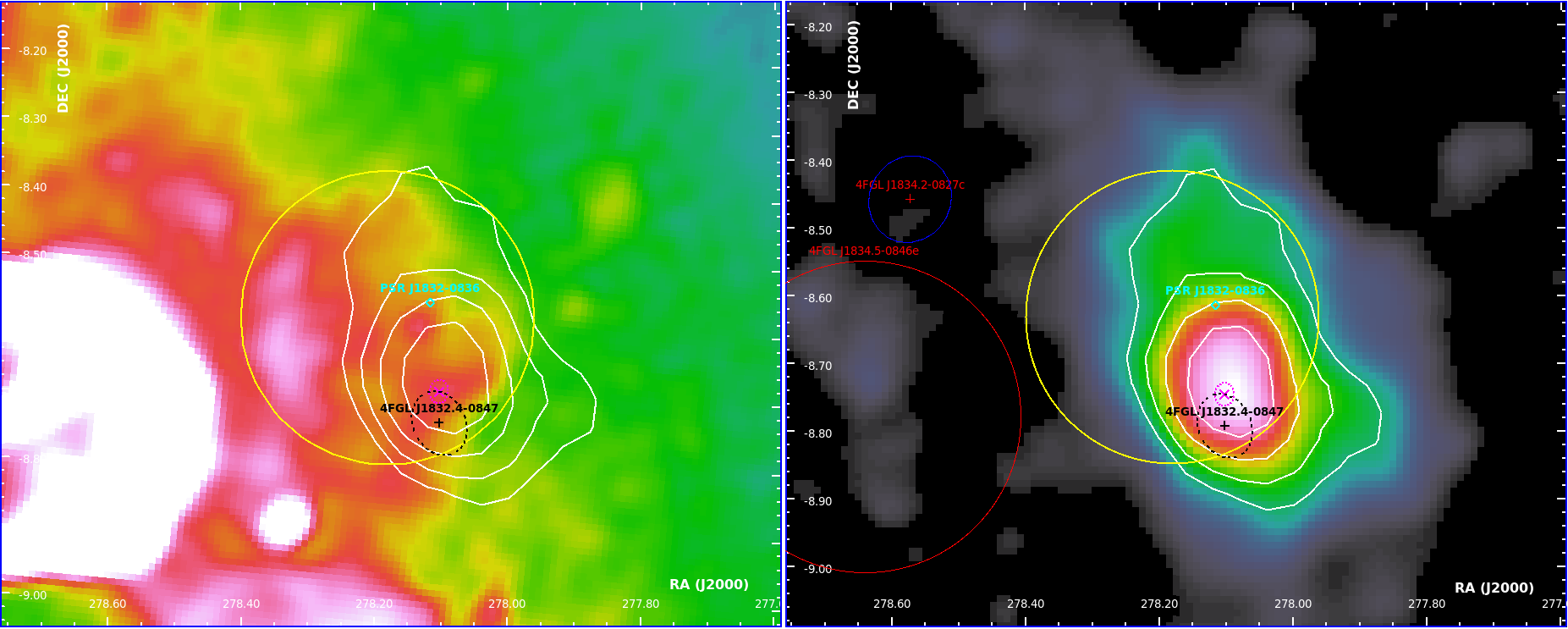}
\vspace{-0.2cm}
\caption{\footnotesize{Radio continuum of the ROI is shown at the left panel and the TS map of {\it Fermi}-LAT data at 1 - 300 GeV is shown at the right panel. The GLEAM location of G23.1$+$0.1 is at the center of the ROI. The white lines are the gamma-ray contours at TS values 25, 36, 49, 64, and 81. The yellow ellipse is the GLEAM extension of G23.1$+$0.1. At the right panel, 4FGL J1832.4$-$0847 is deleted from the background model. Its location is shown with a black cross, while its error ellipse is shown with a black dashed ellipse. The error ellipses of 4FGL sources are for 95\% confidence level. The blue X sign shows the best-fit location and the blue dashed ellipse shows its error. PSR J1832$-$0836, shown with cyan diamond is a pulsar that might have a possible association with this SNR.}}
\vspace{-0.2cm}
\label{figure2}
\end{figure*}
\begin{figure*}
\centering 
\includegraphics[width=0.8\textwidth]{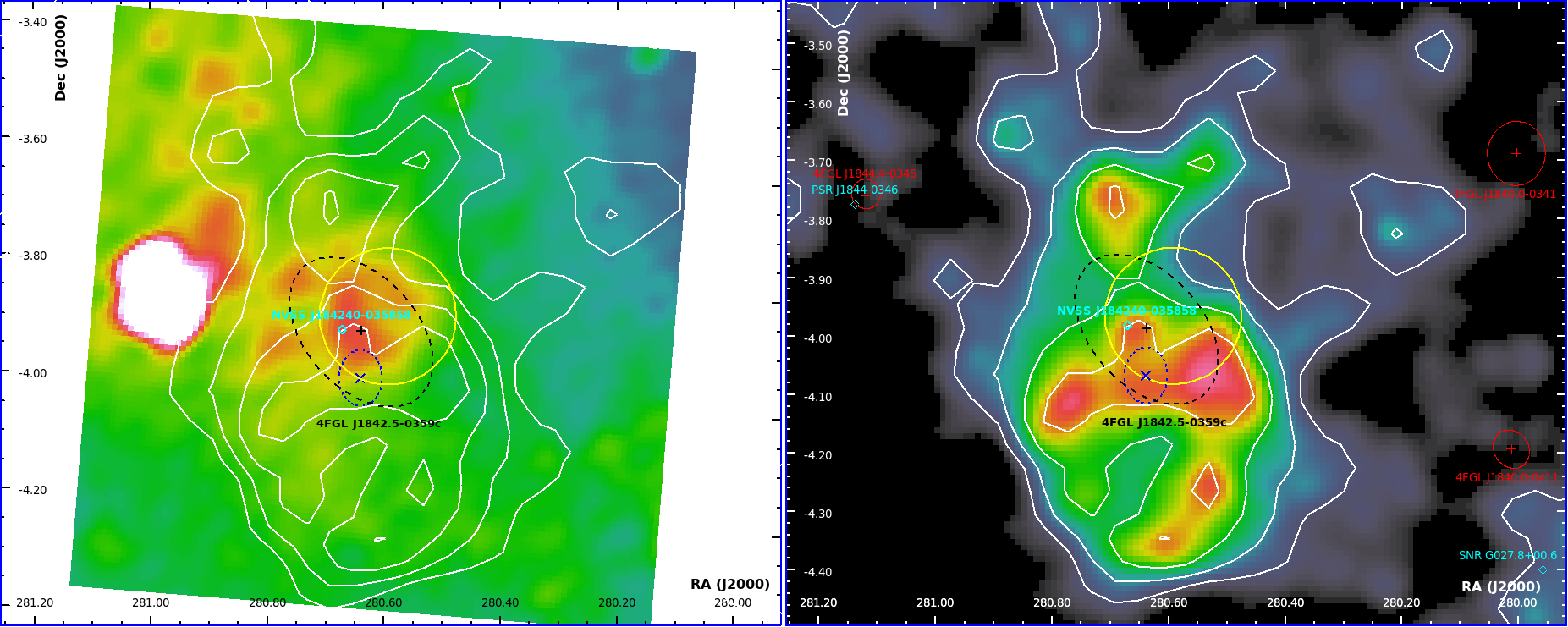}
\vspace{-0.2cm}
\caption{\footnotesize{Radio continuum of the ROI is shown at the left panel and the TS map of {\it Fermi}-LAT data at 1 - 300 GeV is shown at the right panel. The GLEAM location of G28.3$+$0.2 is at the center of the ROI. The white lines are the gamma-ray contours at TS values 4, 9, 16, and 25. The yellow ellipse is the GLEAM extension of G28.3$+$0.2. At the right panel, 4FGL J1842.5$-$0359c is deleted from the background model. Its location is shown with a black cross, while its error ellipse is shown with a black dashed ellipse. Other 4FGL sources are shown with red crosses and their error ellipses are shown with red ellipses. The error ellipses of 4FGL sources are for 95\% confidence level.} The magenta X sign shows the best-fit location and the magenta dashed ellipse shows its error. Other sources from different catalogs are shown with cyan diamonds, where NVSS J184240$-$035858 is an extra-galactic source that is not associated with this SNR.}
\label{figure3}
\end{figure*}

The TS maps produced with the {\it Fermi}-LAT data are shown on the right panels of Figures 1, 2, and 3, for G18.9$-$1.2, G23.1$+$0.1, and G28.3$+$0.2 respectively. Here, it is clearly marked that the error ellipses of the best-fit positions found for each gamma-ray source in this analysis are smaller than the corresponding 4FGL sources given in the {\it Fermi}-LAT catalog. This is due to the fact that we used a bigger data set for each gamma-ray source in comparison to the 8-year {\it Fermi}-LAT catalog's data set. The results of the extension tests show that none of these three SNRs present significant extension with the available {\it Fermi}-LAT data. We obtained 0.16$^{\circ}$, 0.12$^{\circ}$, and 0.27$^{\circ}$ of (Gaussian $\sigma$) extension upper limit values (at 68\%  CL) for G18.9$-$1.2, G23.1$+$0.1, and G28.3$+$0.2, respectively. Therefore, they are detected as point-like gamma-ray sources. This result is consistent with a previous study \citep{Mese2022}, where the detailed extension test results are presented for these three SNRs. At the left panels of Figures 1, 2, and 3, sky maps of GLEAM radio data, which correspond to the same sky regions as the gamma-ray TS maps, are presented. As indicated in different colors, the best-fit position of each detected gamma-ray source is located inside the boundaries of the related GLEAM SNR. Although the results of the extension tests show that G28.3$+$0.2 is best modeled as a point source, a greater structure can be seen in Figure 3, which does not correspond to any known source. We will do further gamma-ray analysis to investigate the excess emission located in the South of the SNR. We also want to find out the nature of the gamma-ray emission, i.e. whether it is hadronic or leptonic. Below is the total photon and energy flux for G18.9$-$1.2, G23.1$+$0.1, and G28.3$+$0.2 in the 0.2 - 300 GeV energy range (with only statistical errors), assuming that the SNRs have a point-like extension and obey a {\it PowerLaw} type spectral model for G23.1$+$0.1 and {\it LogParabola} type model for G18.9$-$1.2 and G28.3$+$0.2: 
\vspace{-0.2cm}
\begin{itemize}
    \item {\bf G18.9$-$1.2}: (4.28 $\pm$ 0.74) $\times$ 10$^{-9}$ cm$^{-2}$ s$^{-1}$ and (5.26 $\pm$ 0.63) $\times$ 10$^{-6}$ MeV cm$^{-2}$ s$^{-1}$
    \vspace{-0.3cm}
    \item {\bf G23.1$+$0.1}: (2.13 $\pm$ 0.22) $\times$ 10$^{-8}$ cm$^{-2}$ s$^{-1}$ and (1.54 $\pm$ 0.13) $\times$ 10$^{-5}$ MeV cm$^{-2}$ s$^{-1}$
    \vspace{-0.3cm}
    \item {\bf G28.3$+$0.2}: (1.41 $\pm$ 0.29) $\times$ 10$^{-8}$ cm$^{-2}$ s$^{-1}$ and (4.75 $\pm$ 0.99) $\times$ 10$^{-6}$ MeV cm$^{-2}$ s$^{-1}$
\end{itemize}
\vspace{-0.2cm}

\begin{figure}
\centering 
\includegraphics[angle=0, width=4.9cm]{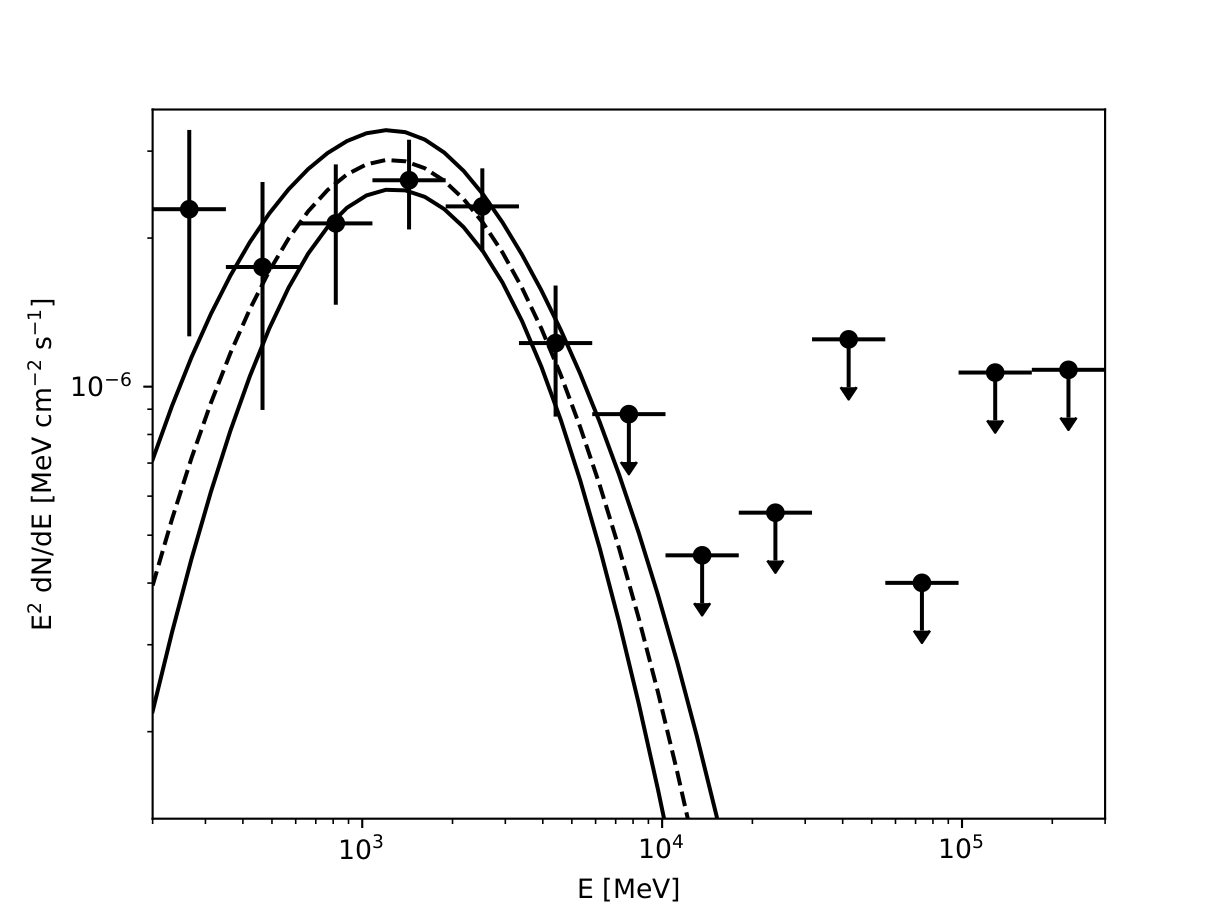}
\includegraphics[angle=0, width=4.9cm]{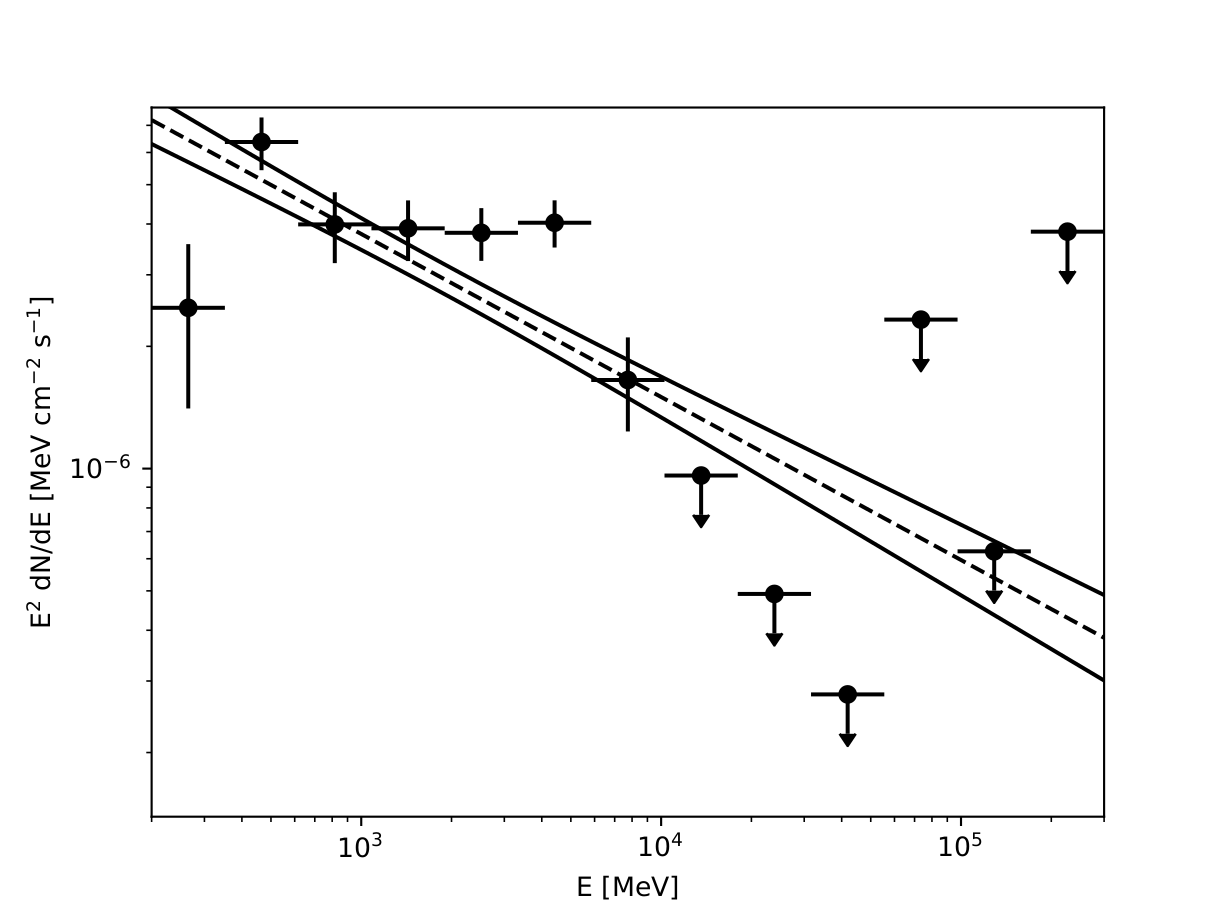}
\includegraphics[angle=0, width=4.9cm]{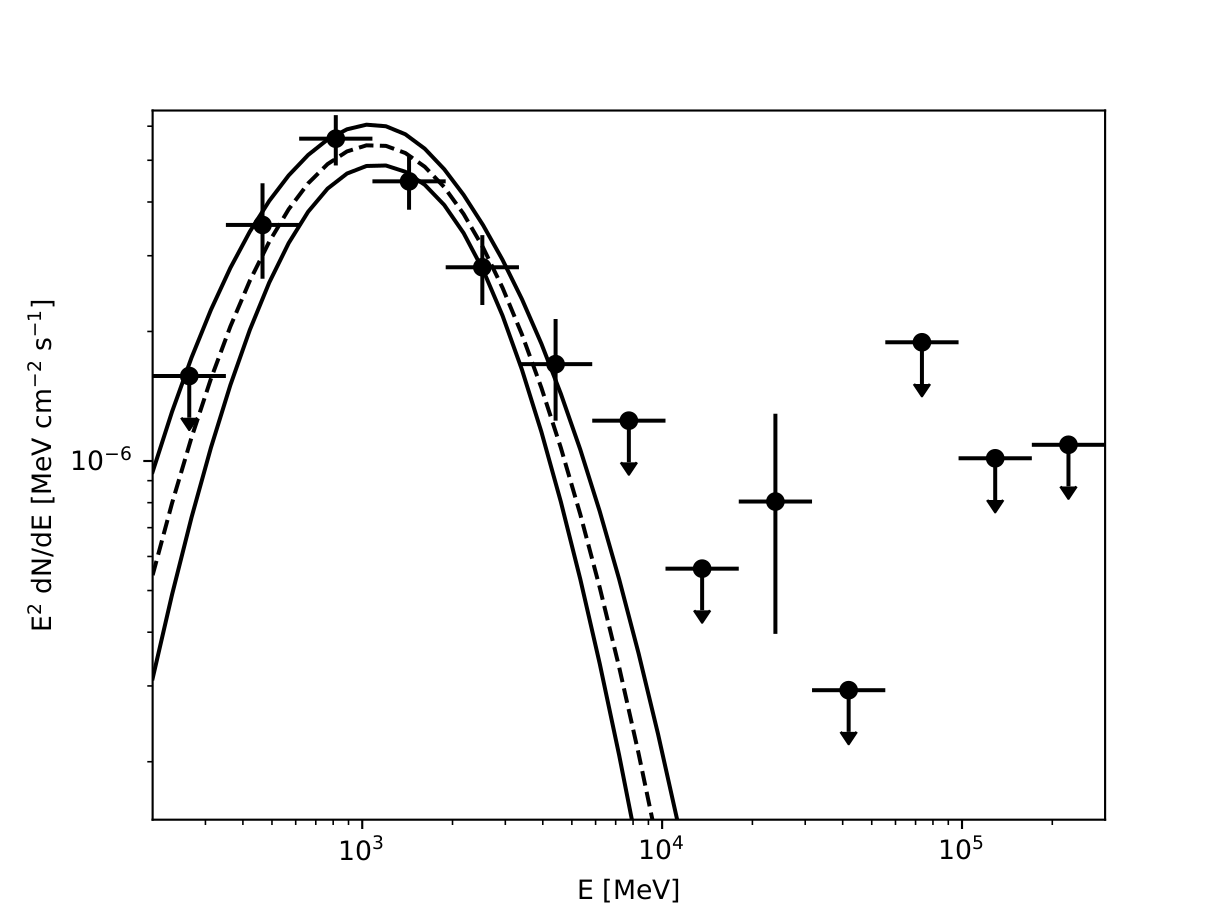}
\vspace{-0.1cm}
\caption{\footnotesize{SED graph of  G18.9$-$1.2, G23.1$+$0.1 and G28.3$+$0.2, at 200 MeV - 300 GeV. The dots represent data points, and vertical lines around the data points show the error bar of the y-axis, where the horizontal lines are for the x-axis. The dotted line is the model fit and the continuous lines represent the error of the model. 
{\it Left panel}: SED graph of G18.9$-$1.2. The spectral model used is {\it LogParabola}.
{\it Middle panel}: SED graph of G23.1$+$0.1, where the {\it PowerLaw} spectral model is used.
{\it Right panel}: SED graph of G28.3$+$0.2. The spectral model used is {\it LogParabola}.}}
\vspace{-0.2cm}
\label{figure4} 
\end{figure}

\vspace{-0.3cm}
\section{Acknowledgments}
\vspace{-0.2cm}
 {\footnotesize We would like to thank the {\it Fermi} collaboration for sharing the {\it Fermi}-LAT data as an open source. Also, we would like to thank Dr. N. H. Walker, for sharing the GLEAM radio data files with us. }

\end{document}